# Transfer Learning and Double U-Net Empowered Wave Propagation Model in Complex Indoor Environments

Ziheng Fu, *Student Member, IEEE*, Swagato Mukherjee, Michael T. Lanagan, Prasenjit Mitra, *Senior Member, IEEE*, Tarun Chawla, and Ram M. Narayanan , *Life Fellow, IEEE*

*Abstract*—**A Machine Learning (ML) network based on transfer learning and transformer networks is applied to wave propagation models for complex indoor settings. This network is designed to predict signal propagation in environments with a variety of objects, effectively simulating the diverse range of furniture typically found in indoor spaces. We propose Attention U-Net with Efficient Networks as the backbone, to process images encoded with the essential information of the indoor environment. The indoor environment is defined by its fundamental structure, such as the arrangement of walls, windows, and doorways, alongside varying configurations of furniture placement. An innovative algorithm is introduced to generate a 3D environment from a 2D floorplan, which is crucial for efficient collection of data for training. The model is evaluated by comparing the predicted signal coverage map with ray tracing (RT) simulations. The prediction results show a root mean square error of less than 3 dB across all tested scenarios, with significant improvements observed when using a Double U-Net structure compared to a single U-Net model.**

## I. INTRODUCTION

THE prediction of signal coverage in complex indoor environments for millimeter-wave communications empowers a variety of exciting 5G indoor solutions, including intelligent healthcare, smart home and manufacturing, cloud services, and low-latency HD livestreaming [1]. An essential component of achieving high network performance required by the Internet of Things (IoT) is accurate coverage prediction in such environments. Hence, we need a generalizable and efficient model that can predict radio coverage with a variety of floorplans and furniture placements.

Previous research efforts proposed stochastic models and developed generalizable and efficient algorithms for estimating channel responses based on the comprehensive analyses of power delay profiles (PDP) and the time-domain distribution of multipath components (MPCs) [2], [3]. However, the analysis on PDP can be challenging in complex scenarios because the necessity for precision in small-scale channel estimation precludes the use of stochastic algorithms.

Furthermore, empirical propagation models for millimeter-wave radio propagation have been extensively studied [4], [5]. These models, featuring adaptive coefficients, address numerous typical indoor scenarios. One subset of these empirical models employs the diffusion approximation of transport theory [6], to depict the exponential decay of wave energy and to compute the excess path loss occurring due to various objects in an indoor environment. However, the performance metrics of these empirical models vary across different sites and are challenged in scenarios with strong multipath effects.

Owing to its deterministic nature, ray tracing (RT) is widely applied in commercial and professional scenarios. RT simulations can be either 2D or 3D and they utilize Geometric Optics (GO) and the Uniform Theory of Diffraction (UTD) to calculate electric fields in a unique virtual representation of the real-world propagation scenarios. However, achieving accurate results using RT in such complex environments necessitates substantial computational resources and time [7]. Also, the need for a unique geometry for each site prevents RT from producing a general radio coverage model.

Machine learning (ML) algorithms can release the limitations of generalizability of RT simulations by employing a large amount of data, producing propagation models with ubiquitous applicability. Bakirtzis et al. [8] proposed a U-Net structure with dilated convolution embedded, and Cisse et al. [9] used a modified conditional Generative Adversarial Networks (cGANs). Moreover, Seretis et al. [11] employed a convolutional neural network (CNN) [10] for indoor propagation, and graph neural network is also an option, as discussed by Liu et al. [12]. These models are interpreting the 2D floorplan into programmable data, incorporating the EM properties of the indoor materials—such as walls, doors, and windows—into maps of permittivity and conductivity as input features. Although the generalizability of the propagation model is insured on the geographical layout of the floorplan, the variability in the configuration of radio transceivers is too diverse to be completely incorporated. More importantly, the complex deployment of furniture exerts a huge influence on the coverage pattern, particularly when operating at high frequencies. Therefore, it is crucial to extend the applicability of ML models in wave propagation by considering the effect of furniture in high-complexity environments, which is the key contribution introduced in this paper.

Critical research in indoor propagation modeling heavily relies on understanding the electromagnetic (EM) response of common indoor materials. Although there is still a need for further research to create a comprehensive database on the effects of building materials on high-frequency radio waves, studies [13], [14], and [15] have provided essential guidelines and quantitative data on the permittivity and conductivity of widely used materials

This paper builds on prior research advancements by generalizing indoor propagation models with furniture. Given the budget of time and computational resources for data collection and model training, we propose a transfer learning and double U-Net based propagation model, specifically designed for complex indoor environments where multiple objects of varied shapes are randomly positioned. Transfer learning leverages a pre-trained model and fine-tunes it on a task-specific dataset, significantly reducing the amount of data and time required for training. Specifically, the problem of propagation modeling is approximated as image translation, where the knowledge acquired by EfficientNet [16], a scalable and highly efficient convolutional neural network architecture, is transferred to tackle our task.

To apply machine learning to indoor propagation modelling, the geographic map is interpreted as images delineating the position and EM properties of objects in the room, which serve as input for the first stage of the double U-Net architecture. Within this stage, a preliminary prediction of coverage is generated. Subsequently, the second stage of the network ingests this coarse prediction before being upsampled, and ultimately yields a detailed coverage prediction. The proposed model takes EfficientNet as the encoder in the U-Net.

The major contributions of this paper are the following:

- A propagation model for furnished indoor environments is proposed, accompanied by a database comprising furniture deployment and their corresponding simulated coverage using RT.
- We developed a MATLAB based tool to construct indoor environments available for performing RT simulations. The database, which consists of processed environments and simulation results, is shared.
- A double U-Net architecture with EfficientNet as a backbone is presented, which is a novel model applicable to multiple image processing tasks and it has demonstrated effectiveness in propagation modelling within complex indoor environments.

The remainder of the paper is organized as follows: Section II and III address the challenges of propagation modeling and the characteristics of furniture blockage in indoor environments. Section IV provides background on the techniques and algorithms that form the core components of the proposed model. In Section V, we discuss the setup of the transceivers and the testing environment. Section VI details the architecture of the proposed model, including the collection and preprocessing of the dataset, followed by the configuration of the training process in Section VII. The analysis of the results are further details are provided in Section VIII.

## II. PROPAGATION MODELLING BY RAY TRACING

In this paper, Remcom's *Wireless Insite©* RT tool is utilized to collect data for training our ML model. RT conducts 3D wave simulations by shooting rays through a defined geometry and utilizes Maxwell's equations to estimate the changes in the field from interactions with the objects in the environment. This section briefly discusses the workflow of RT in *Wireless Insite* and presents the limitations in RT, and therefore introduces the theoretical frameworks that serve as the basis of our work.

To validate the accuracy of RT simulations by *Wireless Insite*, an indoor office environment, as depicted by Maccartney [5, Fig. 2] was replicated in *Wireless Insite*. Using the propagation measurement data for this office provided in Table 19 of [5], the performance of *Wireless Insite* was evaluated by comparing the received power obtained from a simulation conducted at 28 GHz under identical conditions in the paper. The path loss as a function of the transmitter-receiver (T-R) separation distance is presented in Fig. 1. The overall root mean square error (RMSE) between the measurements and simulations is 2.97 dB, with the maximum error reaching 5.1 dB. The correlation coefficient of the two datasets is 0.97. Given the inherent variability in real-world environments, these results provide strong confidence in using *Wireless InSite* as a reliable source of reference data.

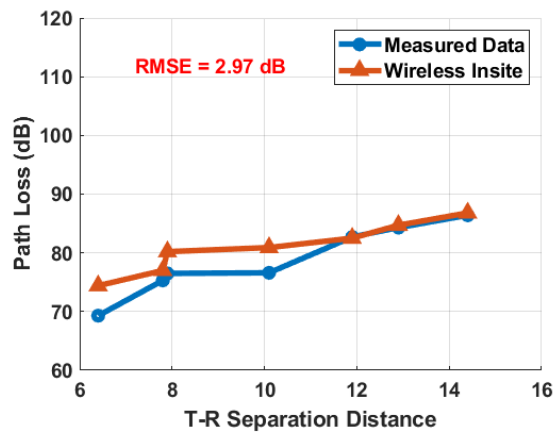

Fig. 1. Received power obtained by measurement and *Wireless Insite*.

### A. Workflow of RT in Wireless Insite

*Wireless Insite* allows for the creation of user-defined indoor environments for signal propagation modelling. The indoor environment can either be imported from CAD files or created using the native Floor Plan Editor in *Wireless Insite*. An example floor plan is shown in Fig. 2(a), where wall heights, ceilings, and windows are specified by the user [17]. Users can place objects with different electrical properties, defined by their relative permittivity, conductivity, and surface roughness, in the indoor environment, as shown in Fig. 2(b). Once the environment is configured, *Wireless Insite* lets the user choose an operating signal frequency, an antenna pattern, and assign these patterns to transmitters (Tx) and receivers (Rx) that can be placed anywhere in the environment.

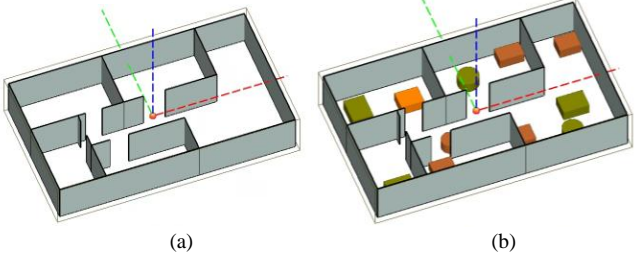

(a)                                (b)

Fig. 2. (a) An unfurnished indoor environment created in *Wireless Insite*, and (b) a furnished one.

After establishing the environment, Tx, and Rx, *Wireless Insite* traces rays through the virtual scene that interact with an object in three ways: when the ray impinges on the face of an object, it can either reflect or transmit, and when it strikes the edge of an object's face, the ray diffracts. The electric field resulting from transmission and reflection off the object is calculated using the Fresnel's Equations [18], while the diffracted field is computed using UTD [19], [20], [21].

Finally, the electric field associated with each ray arriving at the receiver is coherently summed to obtain the received power [17], as shown below:

$$P_R = \frac{\lambda^2}{8\pi\eta_0} \left| \sum_{i=1}^{N_p} E_i \right|^2,$$   (1)

where $N_P$ is the number of paths, $\lambda$ is the wavelength, $\eta_0$ is the characteristic impedance of free space (377 $\Omega$), and $E_i$ is the complex electric field of the $i^{th}$ path at the receiver.

### B. Balance between Computation Time and Accuracy

The computation time is affected by the allowed interaction types and the number of interactions per ray. Reflection interactions extend the path of rays, while transmissions create an additional ray that refracts through a material. Diffractions, however, generate cones of rays from an edge, producing thousands of new rays with each diffraction [17], [22]. This significantly increases the time required for ray tracing time and processing the increased number of rays. On the other hand, the accuracy of RT improves with larger numbers of transmissions ($T$), reflections ($R$), and diffractions ($D$).

To better demonstrate the interplay between accuracy and computation time, we conducted simulations in an indoor environment depicted in Fig. 2. The height of the objects in the scene is 0.76 m (30 inches). The transmitter and receiver grid were positioned slightly above the objects at a height of 0.765 m, and further details about how the objects are generated and placed are provided in Section IV. Fig. 3 presents the cumulative distribution function (CDF) plots of received power with different RT configurations at 5 GHz and 28 GHz, revealing significant differences in the received power distribution. The transmitted power is 0 dBm. Specifically, an improvement of 5–10 dB in received power is observed with more reflections and diffractions, especially at 28 GHz. In other words, simulations that do not account for diffraction or a sufficient number of reflections can result in considerable errors in signal coverage prediction, especially at higher frequencies.

TABLE I
COMPUTATION TIME OF DIFFERENT RT CONFIGURATIONS

| Number of Reflections ($R$) | Number of Transmissions ($T$) | Number of Diffractions ($D$) | Computation Time (s) |
|---|---|---|---|
| 2 | 5 | 0 | 58 |
| 5 | 5 | 0 | 241 |
| 5 | 5 | 1 | 6917 |

However, increasing the number of diffractions and reflections, while enhancing accuracy, also dramatically increases computational costs. Table I presents the computation time for different configurations, with simulations carried out on an NVIDIA GeForce RTX 3060 GPU device, with 28 total cores and 8 CPU threads assigned to computations. While adding 3 reflections (going from 2R5T0D to 5R5T0D) takes 4 times more computation time, the implementation of diffraction (going from 5R5T0D to 5R5T1D) takes even longer and increases the computation time by another factor of 25 times.

### C. Characteristics of Wave Propagation in Furnished Environments

D. Chizhik et al. [23] applied and validated the energy transportation model for wave propagation, which offers insightful discussions on 2D scenarios. However, these scenarios were unfurnished while coverage patterns can vary significantly in furnished indoor environments. Receiver grids that are at or near the height of the furniture in an indoor environment are affected more by furniture blockage and shadowing, while higher receiver grids may show less pronounced shadowing effects.

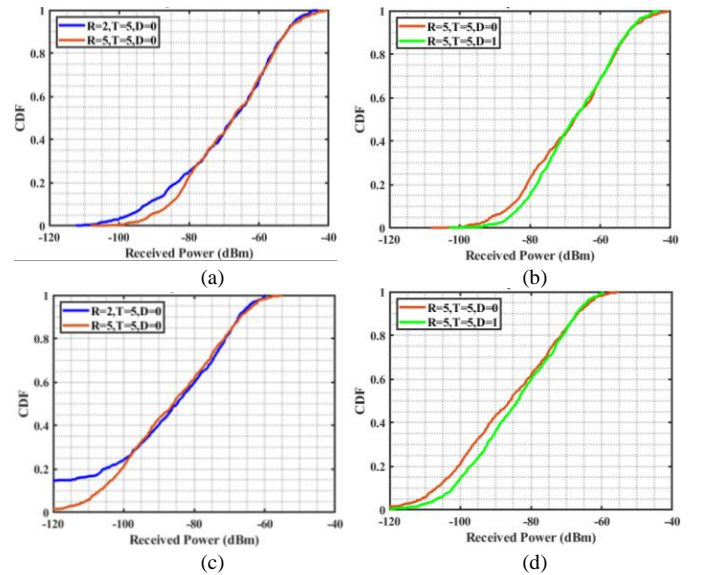

(a)                    (b)

(c)                    (d)

Fig. 3. CDF plots of received power with respect to different RT configurations: (a) $R = 2$, $T = 5$, $D = 0$ and $R = 5$, $T = 5$, $D = 0$ at 5 GHz, (b) $R = 5$, $T = 5$, $D = 0$ and $R = 5$, $T = 5$, $D = 1$ at 5 GHz, (c) $R = 2$, $T = 5$, $D = 0$ and $R = 5$, $T = 5$, $D = 0$ at 28 GHz, (d) $R = 5$, $T = 5$, $D = 0$ and $R = 5$, $T = 5$, $D = 1$ at 28 GHz.

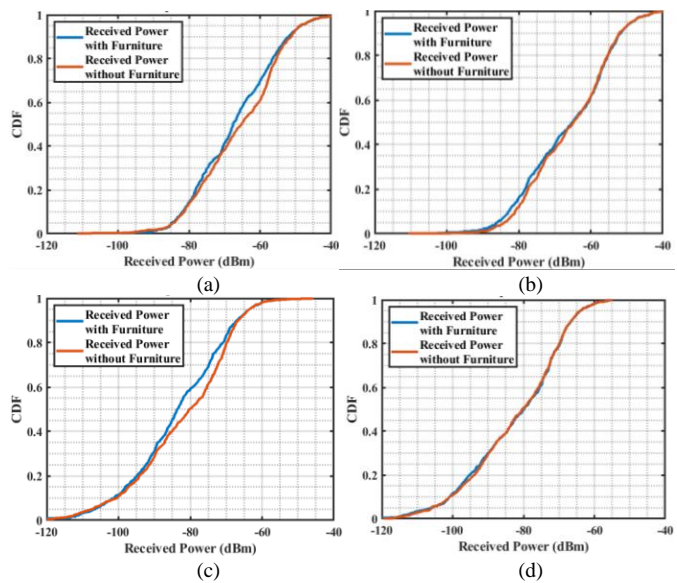

Fig. 4. CDF plots of received power with and without furniture for receiver grid heights of (a) 0.765 m at 5GHz, (b) 1.06 m at 5 GHz, (c) 0.765 m at 28 GHz, and (d) 1.06 m at 28 GHz.

The CDF plots of received power in the indoor environment shown in Fig. 2 are further illustrated in Fig. 4 to investigate the effect of furniture. With the RT configuration of $R = 5$, $T = 5$, and $D = 1$, simulations without furniture for the Rx grid at 0.765 m led to almost 7 dB underestimation of received power at 28 GHz and around 5 dB underestimation at 5 GHz. On the other hand, the Rx grid at a higher height of 1.06 m (40 inches) was relatively unaffected by the presence of the furniture.

In modern, dynamic indoor environments, computational efficiency is not merely a convenience but a necessity, particularly in applications such as augmented reality, indoor navigation, and smart buildings, where real-time feedback is crucial. Traditional ray-tracing algorithms, while accurate, are computationally expensive, time-consuming, and struggle to scale efficiently, especially in large or frequently changing spaces. These methods are impractical for continuous use, dynamic network optimization, or scenarios where users interact with systems that demand immediate responses. In contrast, ML-assisted models offer significant advantages, particularly in resource-constrained environments. Faster predictions enable continuous monitoring and adjustments, essential in settings like factories, hospitals, or large office spaces. By drastically reducing computation time while maintaining acceptable accuracy, the ML model enhances user experience and provides practical, timely solutions, validating the need for speed in modern applications. While a baseline of radio map can be generated by analyzing the basic structure of the floorplan of an indoor environment with machine learning models as discussed in [6], [9], a comprehensive coverage prediction should essentially include the effects of blockage and shadowing by furniture. Since such comprehensive analysis by RT alone needs significant computational resources, we propose a robust ML-based approach that accounts for the effects of furniture.

## III. PROBLEM STATEMENT

Accurately predicting wave propagation patterns in furnished indoor environments is challenging due to the complex interactions between electromagnetic waves and obstacles such as walls, furniture, and other objects. Traditional approaches, such as ray-tracing simulations, often face challenges due to their high computational cost, time-consuming nature, and the need for detailed manual input. On the other hand, empirical models rely on simplified assumptions about the environment, which may not adequately account for the dynamic nature and variability of indoor layouts. The limitation gap in accurate and efficient modeling hinders the optimization of signal coverage, network planning, and overall performance of indoor wireless systems.

To address this, we developed a data-driven ML model to predict wave propagation patterns in furnished indoor environments. The approach begins by encoding furnished floorplans as matrices, which serve as inputs to the ML model. These matrices incorporate geographic information and electromagnetic properties of the objects that constitute the furniture layout. The output of the model will be the radio map corresponding to the specific room layout. Essentially, we aim for the ML model to act as a translator, converting indoor geometry into radio maps.

## IV. RELATED WORK ON NEURAL NETWORKS

This section outlines the foundational concepts and discussions related to the proposed model. We begin with an introduction to CNN, followed by an exploration of the core structures in models that have evolved from CNN, as well as advanced algorithms that significantly enhance the performance of machine learning for complex tasks.

### A. Convolutional Neural Networks

CNN is a class of deep neural networks, highly effective in analyzing visual imagery. Their success in image recognition tasks has demonstrated their power, resulting in widespread adoption and ongoing development [24]. Since an image consists of pixels represented by intensity values in a 2D matrix, image processing is analogous to matrix operations. Consequently, higher-resolution images, which contain more pixels, typically require greater computational power for processing. CNN performs a series of mathematical operations to extract and learn patterns, enabling them to identify objects, textures, and hidden features within an image. A typical CNN consists of several key components as shown in Fig. 5: convolutional layers that apply filters to the input data to create feature maps, activation layers that use an activation function to introduce non-linearity, pooling layers (such as max pooling) that downsample the spatial dimensions to reduce computational load, and fully connected layers that aggregate the features learned by the convolutional layers for final classification.

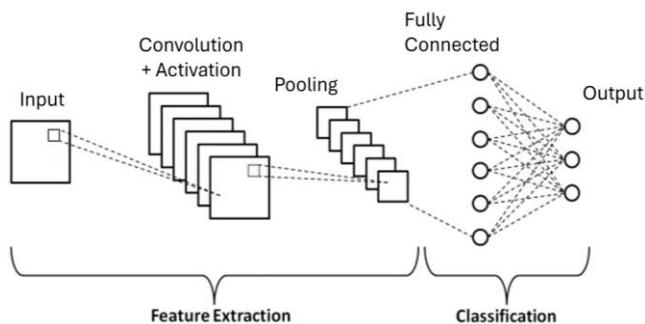

Fig. 5. Outline of CNN structure.

The convolution operation serves as a fundamental block of CNN. The filters or kernels convolve across the input image, computing dot products between the entries of the filter and the input. This process generates feature maps that are essential for recognizing more complex patterns as the network progresses. The stride parameter is pivotal in determining how the filter convolves across the input, a stride of 1 shifts the filter one pixel at a time, while a larger stride moves the filter more pixels per move, resulting in downsampling. Next, the pooling layers reduce the spatial dimensions of the feature representations, which in turn diminishes the number of parameters and the computational burden, enhancing the model's robustness in pattern recognition.

Towards the network's conclusion, fully connected layers undertake the task of classification, leveraging the extracted and downsampled features. Additionally, dropout is employed as a regularization strategy, deactivating a random subset of neurons during the training phase to curb overfitting. Another critical component, batch normalization, adjusts the output from activation layers by normalizing based on the batch mean and standard deviation, thus stabilizing the learning process and significantly lessening the epochs needed for training deep networks.

Despite the effectiveness of traditional CNNs, they face limitations, particularly with very deep architectures. One key issue is vanishing gradients, where gradients become very small during training, causing weights to update slowly, and hindering the learning process. Another issue is overfitting, where the neural network learns the training data too well, capturing noise and details that do not generalize to new, unseen data, leading to poor performance on test data. To address these challenges and enhance the performance on complex visual tasks, advanced architectures such as VGG [24], ResNet [25], Inception network [26], and EfficientNet were developed, each introducing unique elements to overcome the limitations of conventional CNNs. Among these, EfficientNet stands out with a systematic approach to scaling a CNN's depth, width, and resolution. Depth adds layers to learn complex patterns, width expands channels for fine-grained features, and resolution enhances input detail. Using a compound scaling method with fixed coefficients, EfficientNet balances these dimensions for superior performance and fewer parameters. By uniformly scaling width, depth, and resolution with a compound coefficient ($\phi$) and optimized constants ($\alpha$, $\beta$, $\gamma$), it achieves state-of-the-art accuracy in image classification with exceptional efficiency [27].

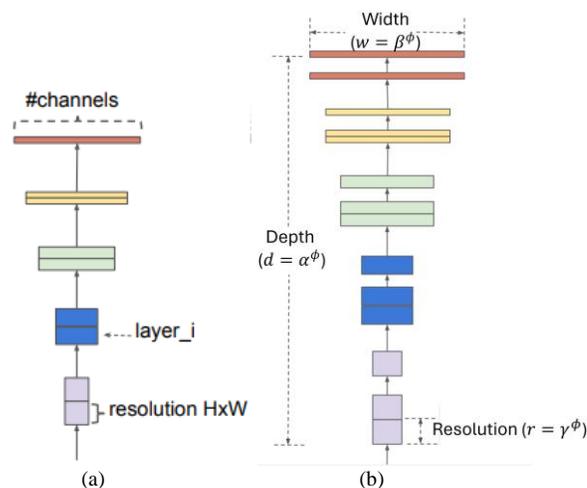

Fig. 6. Structure of (a) baseline neural network, and (b) EfficientNet with compound scaling method that uniformly scales all three dimensions with a fixed ratio [16].

### B. Pre-trained Model and Transfer Learning

The EfficientNet models were initially trained on the ImageNet dataset, a popular benchmark in the field of computer vision, with high-performance GPUs or TPUs to handle the extensive computational requirements. The weights were made available after the training process, as presented by the original author, and can be directly applied to other tasks.

EfficientNet-B0 is the baseline network, developed using a neural architecture search that optimized for both accuracy and efficiency. The subsequent models, B1 to B7, were developed by systematically scaling up B0 using a compound scaling method, where the number in the model's name indicates the depth of the network in terms of the number of layers. Simple models are more computationally efficient than deeper ones, making them suitable for applications where inference speed is critical. On the other hand, deeper variants like EfficientNet-B7 offer better accuracy and are preferable when computational resources are abundant. Hence, these models are excellent candidates for transfer learning,

Tan [16, Fig. 2] presented the performance comparison of various models, where EfficientNet demonstrates substantial competitiveness. This leads us to employ EfficientNet-B5 as the starting point in our task considering the balance between performance and computational resource limitations.

### C. Attention Gate and DoubleU-Net for Image Processing

Evolving from CNN, U-Net was originally designed to efficiently handle biomedical image segmentation in situations where the available data is limited [28]. U-Net extends the traditional CNN architecture by incorporating an encoder-decoder structure as shown in Fig. 7. The encoder, or downsampling path, is similar to a conventional CNN, extracting and condensing features through convolution and pooling operations. This reduces spatial dimensions while increasing feature depth, capturing essential context. The decoder, or upsampling path, mirrors the encoder but performs inverse operations. It uses transposed convolutions and upsampling to reconstruct the image to its original size, enhancing resolution and precision. Skip connections between corresponding encoder and decoder layers preserve spatial information and fine details. The bridge, located between the encoder and decoder, consists of additional convolutional layers that process the compressed features, ensuring relevant information is retained for accurate segmentation. This structure enables U-Net to achieve precise localization and high-resolution outputs. After its initial introduction, U-Net rapidly gained popularity and was adapted and extended for a wide range of other applications beyond the biomedical field [29].

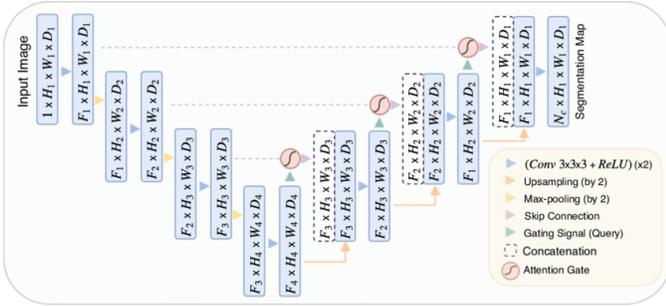

Fig. 7. Structure of a U-Net model with attention gate blocks [30].

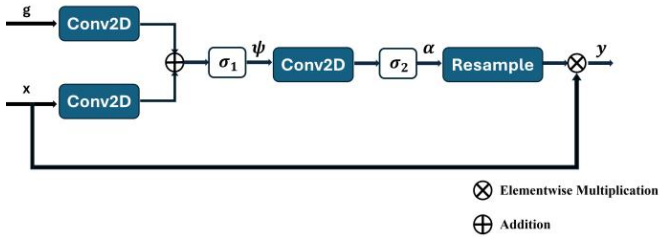

Fig. 8. Structure of attention gate [30].

Building upon the U-Net structure enhanced with dilated convolution layers [6], we have integrated an attention block mechanism [30] into our model. The Attention Gate is aimed at emphasizing only the pertinent regions within an image during the training. The idea of incorporation of attention blocks originated from the hypothesis that radio coverage is influenced by objects within the environment. This methodology allows the model to focus on specific spatial features that are crucial for predicting radio coverage.

As illustrated by Fig. 8, attention blocks in Attention U-Net scale the encoder path's feature maps before concatenating them with the decoder path's corresponding features. This gating mechanism applies a soft mask to prioritize task-relevant features. The gate receives two inputs: the gating signal $g$ from the lower decoder layer, providing rich feature representation, and input features $x$ from the encoder path via skip connections, which retain spatial information.

In addition to the attention gate, DoubleU-Net is used to accelerate the evolution of deep learning models for image processing as it was previously shown to significantly outperform the traditional U-Net [31]. The DoubleU-Net architecture essentially stacks two U-Net models sequentially, with the output of the first U-Net serving as an input to the second. The primary advantage of the DoubleU-Net architecture is its enhanced ability to capture intricate details and improve segmentation accuracy.

## V. CONFIGURATION OF INDOOR ENVIRONMENTS

In this section, we define the indoor environment and detail the configuration of RT simulations. The discussion will also include the collection and pre-processing of data, followed by the construction of the input space for our proposed model.

The data-driven approaches, particularly the deep learning models mentioned in Section III, require a substantial volume of data for effective training. To gather data on coverage patterns in indoor environments where objects are distributed in an arbitrary manner, we developed a script that generates geometric plans based on predefined rules and conducted RT simulations to calculate the path loss.

TABLE II
MATERIAL PROPERTIES [14]

| Material | $\varepsilon_r$ (5 GHz) | $\sigma$ in S/m (5 GHz) | $\varepsilon_r$ (28 GHz) | $\sigma$ in S/m (28 GHz) |
|---|---|---|---|---|
| Concrete | 5.31 | 0.1200 | 5.31 | 0.4838 |
| Wood | 1.99 | 0.0264 | 1.99 | 0.1672 |
| Plasterboard | 2.94 | 0.0362 | 2.94 | 0.1226 |
| Chipboard | 2.58 | 0.0761 | 2.58 | 0.2919 |

The environment is established over a 17 m × 10 m room enclosed by concrete walls. Within this space, various cuboids and cylinders, representing furniture, are distributed randomly. These objects are randomly assigned one of the three common indoor materials: wood, chipboard, or plasterboard. The relative permittivity $\varepsilon_r$ and conductivity $\sigma$ of the materials at the frequency of our interest, 5 GHz and 28 GHz, are presented in Table II.

The objects to be positioned in the room occur in two forms: cuboids with six faces and polygons with 20 faces to approximate cylinders. To replicate practical conditions, the receiver grid height aligns with the typical placement of commonly used indoor devices, such as smartphones, laptops, and tablets, which are generally positioned on desks or cabinets. Accordingly, we set the receiver grid at a height of 0.765 m — 5 mm above typical table height, 0.76 meters (30 inches) — reflecting the fact that antennas in such devices are often mounted within 1 cm from the bottom of their base. An additional receiver grid at a height of 1.06 m (40 inches) represents the scenario of users accessing network services while standing or walking.

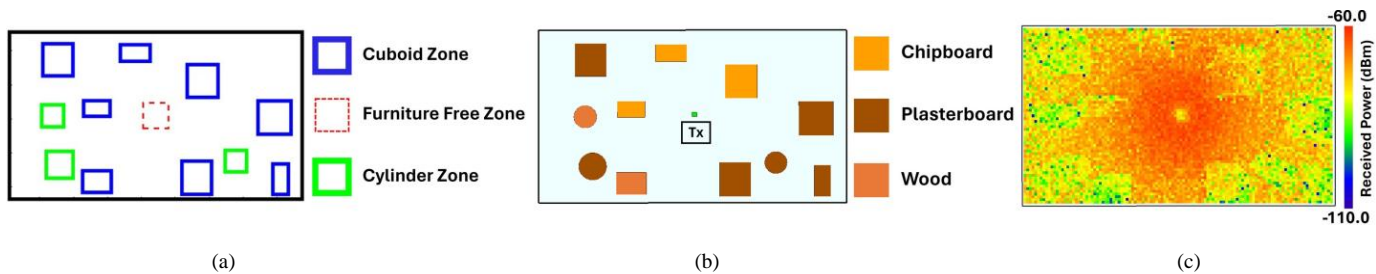

(a)    (b)    (c)

Fig. 9. (a) Geometry plan generated by MATLAB, the cylinder zone is the bounding box to define cylinders, where the third rule of geometry applies; (b) Geometry view of the indoor environment in *Wireless Insite*, the transmitter is placed at the center of the room; (c) Radio map for the geometry shown above, obtained by *Wireless Insite*.

To produce geometry plans that more accurately reflect real indoor environments, we established the following rules:

1) Objects must be randomly placed in the room.
2) To create indoor environments with objects that closely resemble real furniture sizes, the lengths of the cuboids are randomly assigned values between 0.8 and 2.0 meters, while the radii of the cylinders range from 0.5 to 0.8 meters. The cylinders are generated by first defining a bounding box whose edge length is equal to the diameter of the cylinders.
3) To avoid potential overlapping, each pair of objects must have a spacing of at least 0.5 meters.
4) There should be a small area near the transmitter where no object is placed to avoid major signal blockage.
5) The height of each object is 0.76 meters (30 inches).
6) The height of the transmitter is 1.6 meters, positioning it at the typical height of network access points. The height

of two receiver grids are 0.765 meters, which is slightly above the objects, and 1.06 meters (40 inches), respectively.

A MATLAB script is used to generate the geometry shown in Fig. 9(a). The geometry object file is then imported into *Wireless Insite*. A total of 7475 receivers are evenly distributed in the room. Each Rx is assigned an omni-directional antenna pattern, and they are spaced 0.15 m from each other. The Tx is a half-wave dipole antenna with half power beamwidth (HPBW) of 78 degrees mounted at a height of 1.6 m. Both Tx and Rx antennas are vertically polarized. Finally, RT simulations are run with each ray allowed to undergo a maximum of three reflections, three transmissions and one diffraction resulting in the radio map shown in Fig. 9(c).

While the models introduced in Section III were initially designed for image segmentation and classification, their extensive applicability makes them effective methods for predicting indoor propagation loss. The geometry map with the distribution of objects within a space closely resembles an image in its structure as each pixel represents a part of the environment, similar to how pixels in an image represent visual details. By treating the geometry map as an image, where different objects and their layouts are analogous to various features and textures in a picture, a CNN can learn to associate specific configurations with their radio-wave propagation

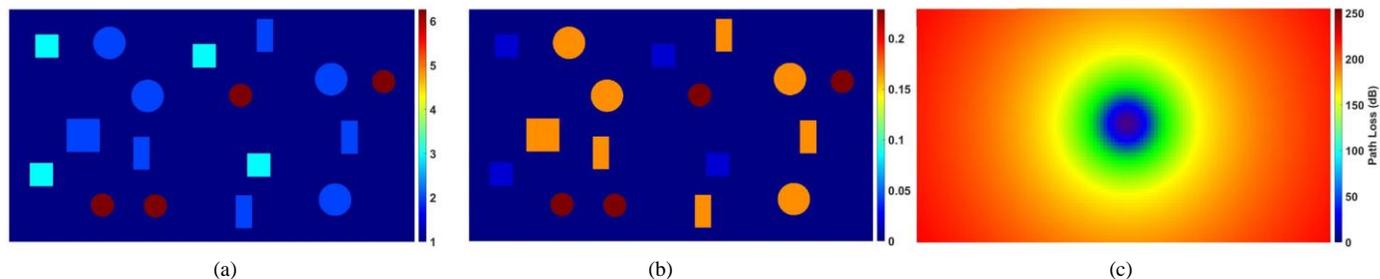

(a)    (b)    (c)

Fig. 10. The input feature with (a) permittivity map, (b) conductivity map, and (c) FSPL, forming the input to the first leg of the network.

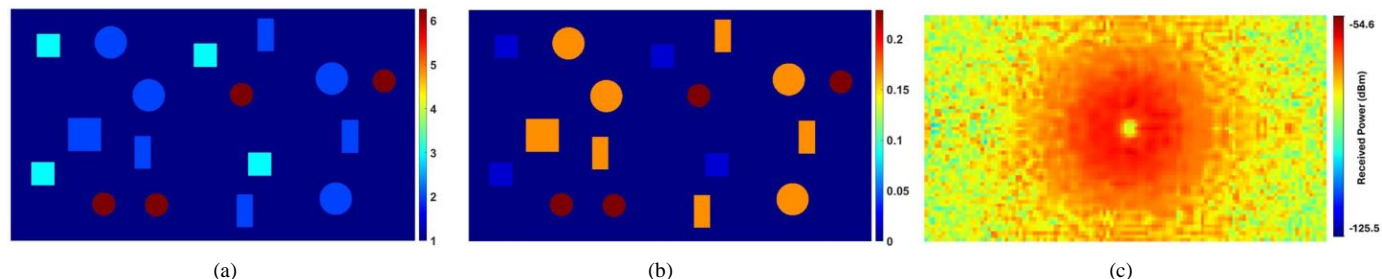

(a)    (b)    (c)

Fig. 11. The input feature with (a) permittivity map, (b) conductivity map, and (c) preliminary prediction, forming the input to the second leg of the network.

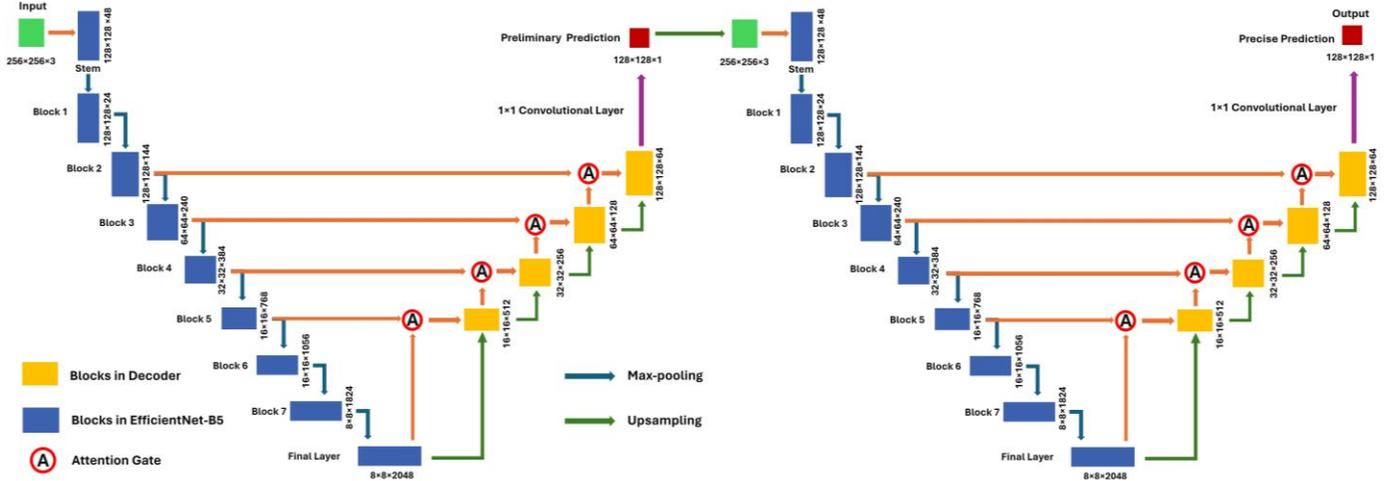

Fig. 12. The architecture of the proposed network.

characteristics.

Up to this point, we have compiled the necessary methodology and theory prior to introducing the model for predicting coverage maps. Building on the groundwork established by previous research, our objective is to create and evaluate a model capable of generalizing radio coverage prediction in complex indoor environments. Our approach involves collecting data and forming input features to feed into the network, followed by assessing the model's generalizability by examining its performance on a testing dataset. This process ensures that our approach is not only data-driven but also robust and applicable to various scenarios, providing a comprehensive assessment of the model's effectiveness in real-world applications.

## VI. PROPOSED METHOD

In this section, we describe the process of converting data obtained from *Wireless Insite* into digital images, which are then used as outputs in the machine learning network. Concurrently, the geometry plan is converted into an input feature that is fed into the network. The details of data processing, and the architecture of the proposed model are elaborated in this section.

### A. Input Features Formation

The input feature has dimensions of 256×256×3. The geometry plan as shown in Fig. 9 is adapted to a permittivity and a conductivity map according to Table II. In regions devoid of objects, the relative permittivity is set to 1 and conductivity is set to 0. The data in the third channel represent the Free Space Path Loss (FSPL), which indicates the transmitter's location and baseline propagation behaviors, in accordance with the Friis transmission equation:

$$P_r^{[dB]} = P_t^{[dB]} + G_t^{[dBi]} + G_r^{[dBi]} + 20\log_{10}\left(\frac{\lambda}{4\pi d}\right) \quad (2)$$

where $P_r$ and $P_t$ are the received and transmitted powers respectively, $G_t$ and $G_r$ are the antenna gains of the transmitting and receiving antennas respectively, $\lambda$ is the wavelength, and $d$ is the distance of the propagation path.

The permittivity and conductivity maps are initially set at dimensions of 1738×997, and the FSPL map has an initial size of 115×65 which equals the size of the receiver grid. To facilitate the training procedure [8] and to prevent memory overload, these maps are resized to 256×256 using bicubic interpolation [32]. The expected output is a 128×128 image with the predicted received power in dBm, which will subsequently be resized to 115×65, which is the original dimension of the receiver grid deployed in the room.

A total of 1000 data samples were gathered through RT simulations using *Wireless Insite*. These samples will subsequently be augmented, and divided into training, validation, and testing sets, with further details provided in Section VI.

### B. Attention DoubleU-Net with EfficientNet as Backbone

Among the models demonstrating brilliant robustness and effectiveness in image translation, we have opted to integrate transfer learning, the Attention mechanism, and the DoubleU-Net architecture for our study. The design of the network we propose is illustrated in Fig. 12, which consists of two U-Nets. Each U-Net comprises a convolutional encoder on the left side and a decoder on the right side. One of the uniqueness of our adapted U-Net exists in the encoder structure, the feature extraction is implemented by an encoder with pre-trained weights from EfficientNet-B5 whose default input image size is 456×456 pixels. However, due to the convolutional nature of the network, it can adapt to different input sizes as well. In our practice, the weights of EfficientNet-B5 are set to be trainable, which allows fine-tuning and enables the network to adjust its learned features to better suit the specifics of our data.

EfficientNetB5 starts with a stem, which is a single convolutional layer. It usually has a kernel size of 3×3 and uses a stride of 2, which reduces the input's spatial dimensions by a factor of 2. In the case of EfficientNet-B5, the stem's convolutional layer has 48 filters, which makes an input image of 456×456×3 halved to 228×228×48.

In the EfficientNet-B5, a sequence of blocks, each comprising multiple layers, follows the initial stem. The stem often performs an initial reduction in dimensionality or complexity of the input data. This can make the computational load more manageable for the deeper layers and can help in focusing on the most relevant features of the input. The blocks incorporate Mobile Inverted Bottleneck Convolution (MBConv) layers, which first increase the number of features, then use a lightweight convolution called depth-wise convolution, and finally reduce the features back. This design helps to reduce computation while maintaining good accuracy, making it suitable for devices with limited processing power. Variations exist among the blocks in terms of kernel sizes, filter counts, and repetition frequencies. From the second block onward, there is a reduction in dimensionality and an increment in depth. Fig. 13 and Fig. 14 illustrate the described architecture and the detailed composition of each module, respectively, with "×3" indicating a sequence of modules within the brackets that are repeated three times.

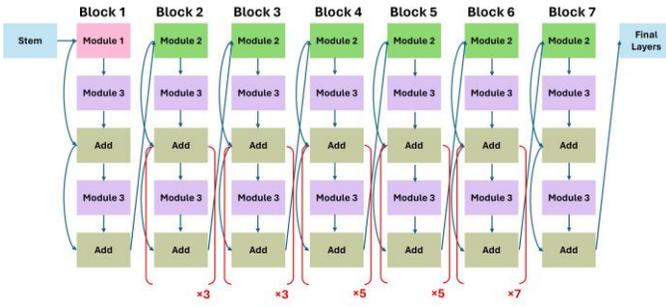

Fig. 13. Architecture of EfficientNet-B5 [32].

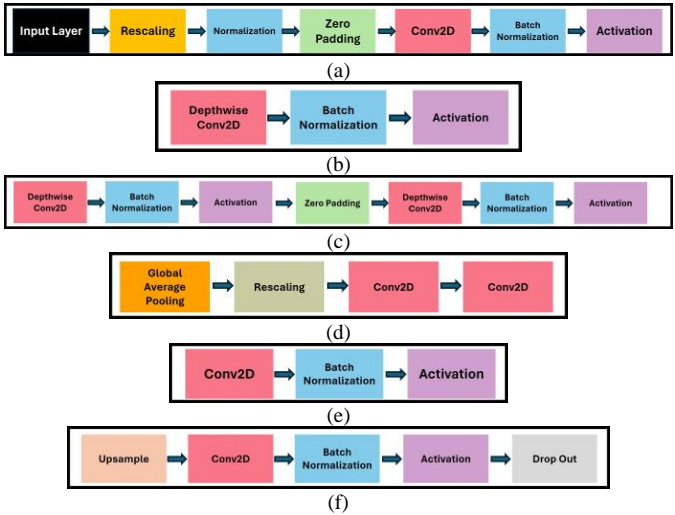

Fig. 14. Architecture of the modules that make up the blocks, (a) is the stem; (b) is Module 1; (c) is Module 2; (d) is Module 3; (e) is the final layer; (f) is the layer in decoders.

In our implementation, Blocks 1, 6, and 7 are excluded from connection to the decoder. This decision is based on the observation that Block 1 predominantly captures basic and low-level features, which are relatively less informative. Conversely, Blocks 6 and 7 are not connected as they are not designed to meet the expected spatial dimensions, while the remaining blocks are sufficient for constructing the model. This approach aims to maintain an optimal balance between the feature richness integrated into the decoder and the computational efficiency of the network. The input image (256×256×3) is zero-padded and then passed through a convolutional layer with a stride of 2 and 48 filters. This operation reduces the spatial dimensions by half and increases the depth to 48. The output size after this layer is 128×128×48. Within the first block (Block 1), the feature map goes through depth-wise convolutions that apply a single filter to each input channel separately, squeeze-and-excitation operations that help the network focus on the most important features by adjusting the importance of different channels, and a projection convolution that reduce the number of channels and reduce the feature map size. These operations maintain the spatial dimensions but transform the depth. By the end of Block 1, the depth is adjusted to 24, maintaining the size at 128×128×24. The output of Block 1 then goes through Block 2, where the output size is changed to 128×128×144 because both blocks use operations with a stride of 1 and same padding in their convolutions. Block 3 and onwards continue to increase the depth while reducing the spatial dimensions. Blocks 1 to 3 primarily extract low-level features, whereas deeper blocks are generally designed to capture high-level and task-specific features. Transferring pre-trained weights for all blocks is not always required and can be selectively adjusted according to specific requirements.

The decoder layers receive information from two distinct paths: one from the deeper layer (the upsampled feature map from the previous decoder layer or, initially, from the bridge layer) and one from the corresponding encoder layer that has passed through an attention gate. The decoder first upsamples the feature map received from the deeper layer, which doubles the spatial dimensions of the feature map to match the dimensions of the corresponding feature map from the encoder path. Then the two feature maps are concatenated. The combined feature map then goes through the convolutional layer within the decoder, and a dropout layer is implemented before it passes to the next decoder layer where the process repeats. A dropout layer randomly sets a proportion of the neurons to 0 at each update during training time, which helps to make the model less sensitive to the specific weights of individual neurons. The dropout rate is set to 0.5 in our implementation. In the final layer, a 1×1 convolution is applied to reduce the channel dimension to the number of output classes.

The output from the first stage of our network is a preliminary estimation of the radio map. This output will undergo nearest neighbor upsampling [33] and will be combined with the permittivity map and conductivity map to serve as input for the second stage of the network where similar processes are executed, as shown in Fig. 12. Ultimately, a precise prediction of the radio map is obtained at the end.

## VII. CONFIGURATIONS OF TRAINING

The effectiveness of training a neural network is influenced not only by its architectural design but also by the proper configuration of network parameters such as learning rate, optimizer and batch size. This section will elaborate on these elements and explore their impact on network training.

### A. Data Augmentation and Workflow

In general, the size of data for training must be large enough for machine learning to yield a robust model. However, the process of data collection can be time consuming with overwhelmingly large amount of RT simulations. Fig. 15 illustrates the training and validation errors corresponding to various sizes of the training set. The intersection of training and validation errors occurs at a training set size of 2100, which suggests a balance point to manage computational costs and to avoid overtraining. Considering the inherent variability in performance during training, we initially collected 1000 data samples and subsequently expanded the dataset to 3000 samples through augmentation.

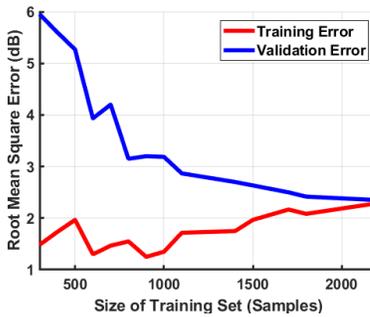

Fig. 15. Training and validation error with respect to the size of training set.

The dataset, comprising 1000 data samples obtained from *Wireless Insite*, can be expanded by modifying the geometry, FSPL, and radio map. This is achieved by flipping these elements upside down and rotating them from left to right, which triples the size of data. The augmented dataset is shuffled before it is divided into three subsets: 2000 samples for training, 500 for validation, and 500 for testing. The training data will be input into the first stage of the network, yielding 2000 preliminary results. Following this, the network parameters will be confirmed after the second stage is trained using the preliminary predictions. Finally, a testing set will be employed to evaluate the overall performance of the network.

### B. Model Training Parameters

In the training of our model, we utilized the Adaptive Moment Estimation (Adam) optimization algorithm [34], a widely employed method in machine learning, particularly for deep learning neural networks. It adaptively adjusts the learning rates of each parameter based on estimates of first and second moments of the gradients, promoting faster and more stable convergence in training deep learning models. The learning rate is initially set to $1e^{-2}$ and it decreases linearly to $1e^{-8}$. The batch size was set to 32, indicating that 32 samples were processed by the network in each iteration and the number of iterations per epoch is the number of training samples divided by batch size. The training was conducted over 30 epochs with an early stopping criterion based on the cessation of decrease in validation and training errors over the last three epochs. The two stages are trained sequentially as the second stage is trained following the completion of the first stage's training. The model is trained with Nvidia A100 GPU supported by Google Colab, and the code and data used for this study are available for download at https://github.com/funited/indoor-coverage-prediction.

### C. Evaluation Metrics

In addition to employing a dropout layer, we introduced Ridge regularization, also known as $l_2$ regularization, during the training process. This technique adds a penalty, $P$, to the loss function based on the magnitude of weights in the network. The penalty is simply the sum of the squares of all the feature weights multiplied by the regularization factor $\xi$, which controls the extent of regularization. A higher value means more regularization, pushing the model towards simpler models to avoid overfitting. Conversely, an excessive value can lead to underfitting, where the model is overly simplified and fails to capture the underlying pattern. In our simulations, we applied $l_2$ regularization with a $\xi$ value of 0.01. The penalty is given by

$$P = \xi \sum_{k=1}^{K} w_k^2 \tag{3}$$

where $K$ is the total number of weights in the model and $w_k$ is the unique set of weights of each filter in the convolutional layers.

The performance of the network is evaluated by root mean square error (RMSE), which is commonly used for regression problems. The RMSE computes the square root of the average of the squares of the differences between the predicted values and the actual values. With the penalty term, the loss function $L$ can be expressed as:

$$L = \sqrt{\frac{1}{NA} \sum_{i=1}^{NA} (y_i - \hat{y}_i)^2} + \xi \sum_{k=1}^{K} w_k^2 \tag{4}$$

where $N$ is the number of training samples, $A$ is the total number of receivers, $y_i$ is the true received power for the $i^{th}$ receiver, $\hat{y}_i$ is the predicted received power for the $i^{th}$ receiver, $w_k$ is the weight of the $k^{th}$ feature, and $K$ is the total number of features of the network. The mean absolute percentage error (MAPE) and Pearson correlation coefficient are also commonly applied metrics for model predictions. MAPE is a metric used to measure the accuracy of predictions in comparison to actual values. It is expressed as a percentage and represents the average magnitude of errors between predicted values and actual values, relative to the actual values. The formula for MAPE is:

$$\text{MAPE} = \frac{1}{n} \sum_{i=1}^{n} \left| \frac{y_i - \hat{y}_i}{y_i} \right| \times 100\% \tag{5}$$

where $N$ is the number of observations, $y_i$ is the actual value while $\hat{y}_i$ is the predicted value for observation.

The Pearson correlation coefficient, on the other hand, quantifies the linear relationship between two variables. It indicates how strongly two variables are related and whether they move in the same direction (positive correlation) or opposite directions (negative correlation). The Pearson correlation is calculated as:

$$r = \frac{\text{cov}(X,Y)}{\sigma_X \sigma_Y} \tag{6}$$

In addition, the structural similarity index (SSIM) [35], a measurement of the visual similarity between two images by considering changes in luminance, contrast, and structure, is applied to evaluate the radio map predictions by the trained model. A SSIM score closer to 1 generally indicates superior model performance.

## VIII. RESULTS OF COVERAGE PREDICTION

This section presents both statistical and visual results to examine the generalizability and robustness of the proposed model. The model is assessed over four scenarios described in Table III with different frequency bands and different receiver grid heights. Table IV presents the performance evaluation over the four scenarios. The model yields an overall testing error below 3 dB in all scenarios. The overall testing error is computed by RMSE of received power across the receiver grid of all testing samples. The mean absolute error (MAE) for each scenario is also listed. The MAPE, Pearson correlation coefficients, and SSIM scores across the samples for each scenario are presented in Table V.

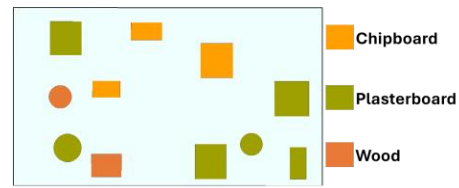

Fig. 16. An exemplar floorplan.

TABLE III
CONFIGURATIONS OF THREE SCENARIOS

| Scenario | $f$ (GHz) | Rx Height (m) |
|---|---|---|
| 1 | 5 | 0.765 |
| 2 | 5 | 1.060 |
| 3 | 28 | 0.765 |
| 4 | 28 | 1.060 |

### A. Performance Evaluation

A floorplan depicted by Fig. 16 is selected as the indoor environment for presenting the results, and it is generated by the methodology described in Section IV. Figs. 17–20 highlight the predicted and actual signal strength distributions within the indoor environment in Fig. 16, where warmer colors represent higher power levels. The enhanced coverage map predictions, using preliminary results from the first stage as a reference, can confirm that the two-stage architecture design is effective. This progress is demonstrated by Table VI, which presents the RMSE values for both the first stage (S1) and the second stage (S2).

The heatmaps in Fig. 17 and Fig. 19 show that objects in indoor environments significantly influence propagation patterns. The largest RMSE observed in the test samples is 2.26 dB, occurring under scenario 3. It is also observed that signal attenuation is more pronounced at 28 GHz.

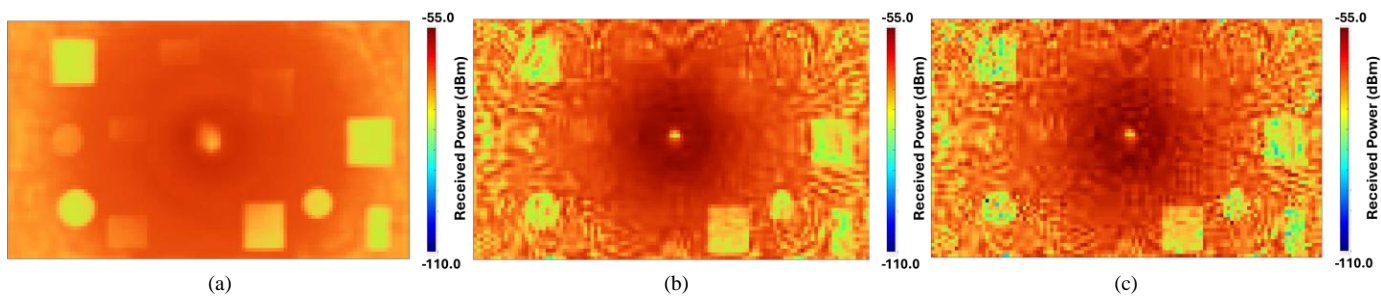

Fig. 17. Radio maps obtained by (a) preliminary prediction at the first stage, (b) final prediction at the output, and (c) RT simulations of Wireless Insite at 5 GHz and receiver grid at 0.765 m.

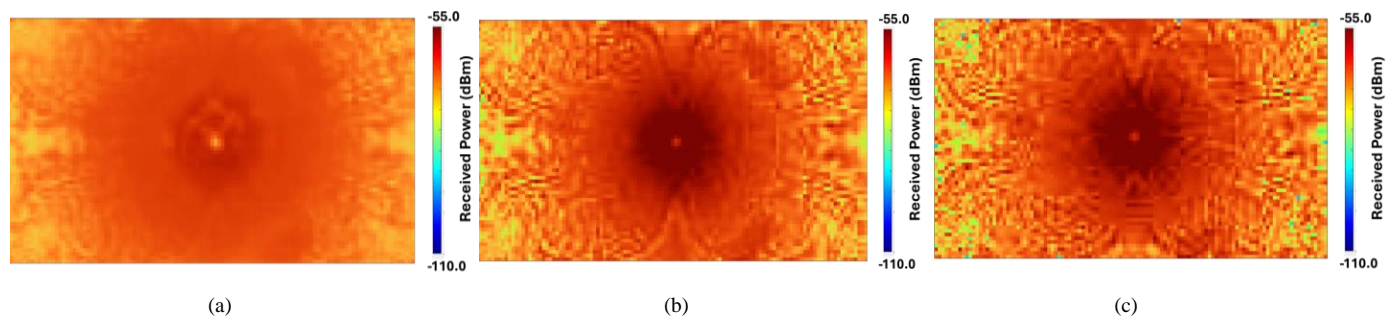

Fig. 18. Radio maps obtained by (a) preliminary prediction at the first stage, (b) final prediction at the output, and (c) RT simulations of Wireless Insite at 5 GHz and receiver grid at 1.06 m.

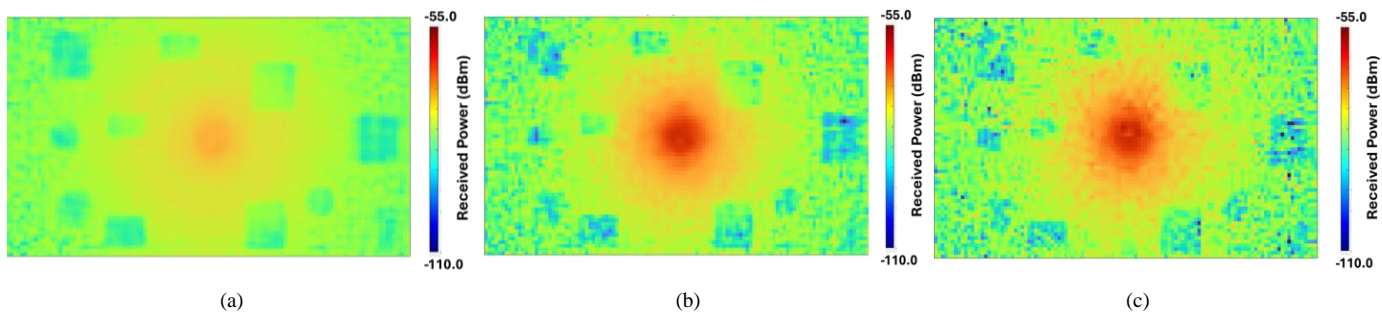

Fig. 19. Radio maps obtained by (a) preliminary prediction at the first stage, (b) final prediction at the output, and (c) RT simulations of Wireless Insite at 28 GHz and receiver grid at 0.765 m.

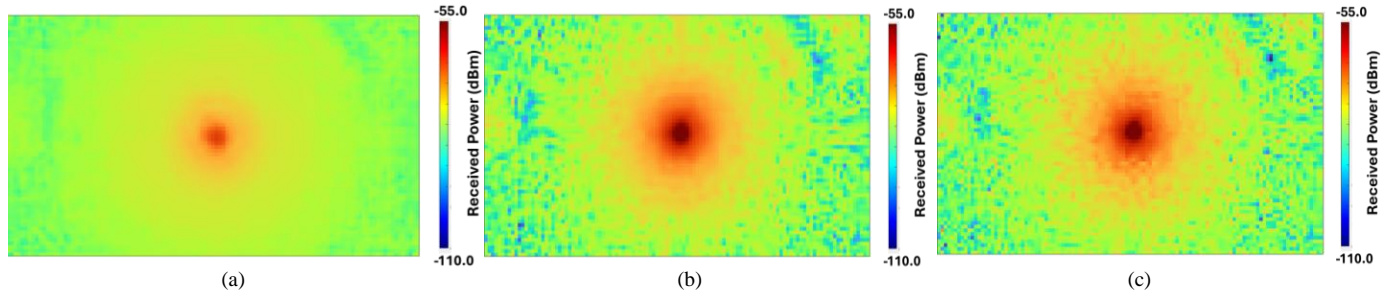

Fig. 20. Radio maps obtained by (a) preliminary prediction at the first stage, (b) final prediction at the output, and (c) RT simulations of Wireless Insite at 28 GHz and receiver grid at 1.06 m.

The plots of cumulative distribution function (CDF) of the root of square error between the predicted and simulated radio maps are provided in Fig. 21, demonstrating that a significant portion of predictions closely aligns with the simulations, with most errors concentrated below a threshold, thus confirming the accuracy and reliability of the proposed predictive model in complex indoor scenarios.

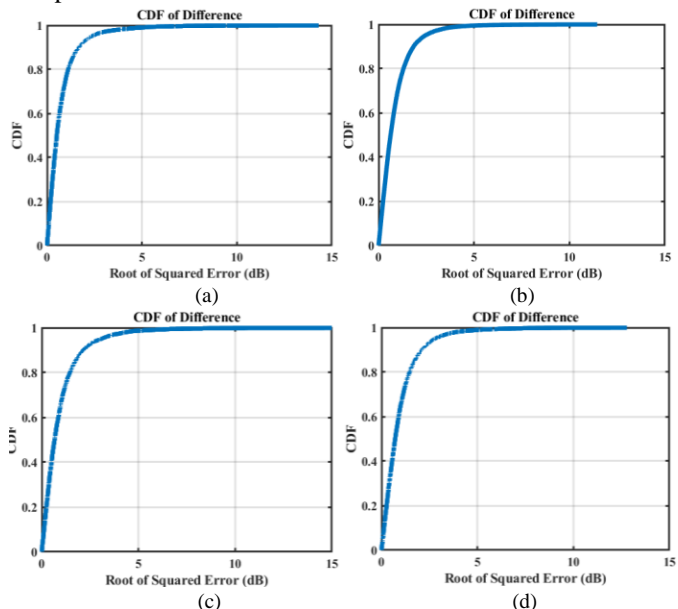

Fig. 21. CDF plots of the root of squared error between the coverage of prediction and RT simulations with respect to the example shown in (a) Fig. 17, (b) Fig. 18, (c) Fig. 19, and (d) Fig. 20.

TABLE IV
PERFORMANCE ANALYSIS OF DIFFERENT SCENARIOS

| Scenario | Training Error (dB) | | Validation Error (dB) | | Testing Error (dB) | |
|---|---|---|---|---|---|---|
| | RMSE | MAE | RMSE | MAE | RMSE | MAE |
| 1 | 2.18 | 0.97 | 2.60 | 1.31 | 1.92 | 1.19 |
| 2 | 2.10 | 0.93 | 2.71 | 1.38 | 2.06 | 1.25 |
| 3 | 2.36 | 1.20 | 2.69 | 1.62 | 2.26 | 1.41 |
| 4 | 2.24 | 1.16 | 2.60 | 1.56 | 2.17 | 1.38 |

TABLE V
MAPE, SSIM SCORES AND CORRELATION COEFFICIENTS ($r$) FOR TRAINING, VALIDATION, AND TESTING SAMPLES

| Scenario | Training | | Validation | | Testing | |
|---|---|---|---|---|---|---|
| 1 | MAPE | 1.60% | MAPE | 2.15% | MAPE | 1.96% |
| | SSIM | 0.96 | SSIM | 0.84 | SSIM | 0.83 |
| | $r$ | 0.95 | $r$ | 0.90 | $r$ | 0.92 |
| 2 | MAPE | 1.57% | MAPE | 2.30% | MAPE | 2.10% |
| | SSIM | 0.99 | SSIM | 0.85 | SSIM | 0.92 |
| | $r$ | 0.97 | $r$ | 0.90 | $r$ | 0.92 |
| 3 | MAPE | 1.48% | MAPE | 1.89% | MAPE | 1.87% |
| | SSIM | 0.96 | SSIM | 0.87 | SSIM | 0.87 |
| | $r$ | 0.95 | $r$ | 0.92 | $r$ | 0.92 |
| 4 | MAPE | 1.46% | MAPE | 1.89% | MAPE | 1.86% |
| | SSIM | 0.98 | SSIM | 0.89 | SSIM | 0.92 |
| | $r$ | 0.96 | $r$ | 0.92 | $r$ | 0.92 |

TABLE VI
RMSE FOR FIRST STAGE (S1) AND SECOND STAGE (S2)

| Scenario | Training Error (dB) | | Validation Error (dB) | | Testing Error (dB) | |
|---|---|---|---|---|---|---|
| | S1 | S2 | S1 | S2 | S1 | S2 |
| 1 | 5.18 | 2.18 | 5.79 | 2.60 | 5.61 | 1.92 |
| 2 | 4.80 | 2.10 | 6.47 | 2.71 | 5.51 | 2.06 |
| 3 | 4.92 | 2.36 | 6.26 | 2.69 | 4.80 | 2.26 |
| 4 | 5.44 | 2.24 | 6.27 | 2.60 | 5.09 | 2.17 |

### B. Verifications on Transfer Learning

A prominent improvement in prediction performance can be observed when comparing models trained from random initialization to the one utilizing transfer learning with fine-tuning. The proposed model exploits EfficientNet-B5 with the weights pre-trained on ImageNet which contains over 14 million labeled images across thousands of categories. With the performance of the proposed model demonstrated in this section, a similar training process was conducted using the same model as shown in Fig. 12 but initialized entirely with random weights (scratch training). As shown in Fig. 22, the proposed model, assisted by transfer learning, achieves better accuracy in radio map predictions under the same conditions of labeled data and computational resources, and the model converges within fewer training epochs.

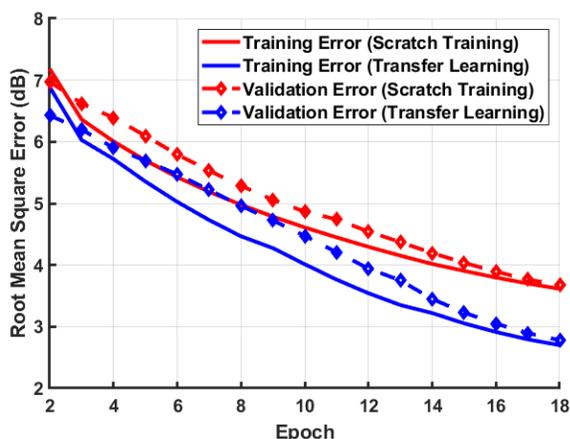

Fig. 22. Training performance with and without transfer learning, from epoch 2 to 18.

### C. Performance Comparison with Large-Scale Path Loss Models

Existing machine learning models have not yet focused on furnished indoor environments, making it unsuitable to directly compare their performance with the proposed model. Instead, large-scale path loss models specifically designed for non-line-of-sight (NLOS) scenarios in indoor hotspot (InH) office environments [4] are the most appropriate for the type of indoor environment considered in this study. The large-scale models, including the Close-In (CI) path loss model and the Floating Intercept (FI) path loss model (ABG), are used to calculate the path loss across the randomly designed floorplans. As we applied these models on the 1000 floorplans that we generated, the average RMSE value is employed as the metric for comparison with the proposed model.

Table VIII shows that the proposed model significantly outperforms the large-scale path loss models, achieving the lowest RMSE values across all scenarios and thus, underscores a significant advancement in prediction accuracy for furnished indoor propagation scenarios.

TABLE VII
PERFORMANCE COMPARISON AMONG LARGE-SCALE PATH LOSS MODELS
AND THE PROPOSED MODEL

| Model | Average RMSE across floorplans for each scenario (dB) | | | |
|---|---|---|---|---|
| | 1 | 2 | 3 | 4 |
| 3GPP TR 38.901 Indoor-Office NLOS model | 10.62 | 11.00 | 10.89 | 11.47 |
| 5GCM InH Indoor-Office NLOS single slope CIF model | 9.62 | 9.98 | 11.09 | 11.68 |
| 5GCM InH Indoor-Office NLOS single slope ABG model | 6.42 | 6.14 | 8.29 | 8.23 |
| 5GCM InH Indoor-Office NLOS dual slope ABG model | 6.71 | 6.08 | 6.97 | 7.19 |
| mmMAGIC InH NLOS model | 5.87 | 5.38 | 7.92 | 8.35 |
| Proposed model | 1.92 | 2.06 | 2.26 | 2.17 |

## IX. CONCLUSION

This article systematically explores the challenges and solutions associated with predicting radio coverage in complex indoor environments for millimeter-wave communications. A significant focus is placed on mitigating issues like radio dead zones by sophisticated modeling techniques, such as RT simulations and advanced machine learning algorithms, including U-Nets enhanced with EfficientNet as the backbone. These methods are tailored to accommodate the random layouts and object placements typical of indoor settings. The introduction of a novel machine learning architecture, combining transfer learning with a double U-Net structure, marks a substantial advancement in the field. This model leverages the power of EfficientNet and attention mechanisms to enhance generalizability and accuracy in predicting indoor radio coverage. The empirical evaluations demonstrate that this model effectively reduces prediction error and adapts to various furnished indoor scenarios, reflecting its robustness and efficiency. As the wave propagation in indoor environments is primarily influenced by two major factors—the layout of walls and the presence of furniture—it is important to note that this model can be integrated with previous work on radio map predictions in unfurnished environments to provide a comprehensive solution.

In conclusion, the research contributes a significant step forward in the predictive modeling of indoor radio coverage for 5G and beyond, offering a promising tool for enhancing connectivity in diverse indoor scenarios. Future work can further refine these models, perhaps by integrating even more detailed environmental data or by expanding the model's adaptability to different frequency bands and more complex geometrical challenges.